\documentclass[preprint,amsmath,amssymb,aps,eqsecnum,nofootinbib]{revtex4}
\usepackage{bm}
\usepackage{mathrsfs}
\usepackage[dvipdfmx]{graphicx}

\begin{document}
\title{Revisiting the Aretakis constants and instability in two-dimensional Anti-de Sitter spacetimes}
\author{Takuya Katagiri}
\author{Masashi Kimura}
\affiliation{Department of Physics, Rikkyo University, Toshima, Tokyo 171-8501, Japan}
\date{\today}
\preprint{RUP-21-2}

\begin{abstract}
We discuss dynamics of massive Klein-Gordon fields in two-dimensional Anti-de Sitter spacetimes ($AdS_2$), in particular conserved quantities and non-modal instability on the future Poincar\'e horizon called, respectively, the Aretakis constants and the Aretakis instability. We find out the geometrical meaning of the Aretakis constants and instability in a parallel-transported frame along a null geodesic, i.e., some components of the higher-order covariant derivatives of the field in the parallel-transported frame are constant or unbounded at the late time, respectively. Because $AdS_2$ is maximally symmetric, any null hypersurfaces have the same geometrical properties.  Thus, if we prepare parallel-transported frames along any null hypersurfaces, we can show that the same instability emerges not only on the future Poincar\'e horizon but also on any null hypersurfaces. This implies that the Aretakis instability in $AdS_2$ is the result of singular behaviors of the higher-order covariant derivatives of the fields on the whole $AdS$ infinity, rather than a blow-up on a specific null hypersurface. It is also discussed that the Aretakis constants and instability are related to the conformal Killing tensors. We further explicitly demonstrate that the Aretakis constants can be derived from ladder operators constructed from the spacetime conformal symmetry.
\end{abstract}

\maketitle

\section{Introduction}
The Anti-de Sitter spacetime is a maximally symmetric spacetime with negative constant curvature and a unique solution, which is strictly stationary, of Einstein's equations with a negative cosmological constant \cite{Boucher:1983cv}. In theoretical physics, the Anti-de Sitter spacetime has a central role because of the $AdS$/CFT correspondence \cite{Maldacena:1997re}. The timelike conformal infinity makes it fail to be globally hyperbolic, and the negative cosmological constant realizes a confined system. Due to that, non-trivial phenomena are induced, e.g., turbulent instabilities, superradiant instabilities, and holographic superconductors \cite{Cardoso:2004hs,Ishibashi:2004wx,Bizon:2011gg,Bizon:2017yrh,Hartnoll:2008kx}. Those motivate us to study dynamics of fields in the (asymptotically) Anti-de Sitter spacetime. The Anti-de Sitter spacetime has also been studied from the point of view of the near-horizon geometry~\cite{Kunduri:2013gce}, i.e., two-dimensional Anti-de Sitter spacetime ($AdS_2$) structures appear in the vicinity of extremal black hole horizons. Thus, it is expected that the study of $AdS_2$ brings us insight into fundamental properties near the horizon of the extremal black holes. 

Aretakis has shown that the (higher-order) derivatives of test massive scalar fields blow up at the late time along the event horizon in the four-dimensional extremal Reissner-Nordstr\"om black holes \cite{Aretakis:2011ha,Aretakis:2011hc}, which is called the Aretakis instability. In ~Refs. \cite{Aretakis:2012ei, Lucietti:2012xr,Lucietti:2012sf,Murata:2012ct,Gralla:2019isj}, it has also been shown that the same phenomena occur in other extremal black hole spacetimes and for other fields. Many aspects of the Aretakis instability have been studied in~Refs. \cite{Bizon:2012we,Aretakis:2012bm,Aretakis:2013dpa,Murata:2013daa,Angelopoulos:2016wcv,Zimmerman:2016qtn,Angelopoulos:2018yvt,Godazgar:2017igz,Gralla:2017lto,Gralla:2018xzo,Angelopoulos:2018uwb,Bhattacharjee:2018pqb,Cvetic:2018gss,Angelopoulos:2019gjn,Burko:2020wzq,Hadar:2017ven,Hadar:2018izi}. These suggest that the Aretakis instability is the robust phenomena around the extremal black holes. Thus, it is interesting to study the Aretakis instability from the point of view of the near-horizon geometry~\cite{Lucietti:2012xr,Zimmerman:2016qtn,Godazgar:2017igz,Gralla:2017lto,Gralla:2018xzo,Hadar:2017ven,Hadar:2018izi}. 

The Aretakis instability of massive scalar fields in $AdS_2$ has already been discussed \cite{Lucietti:2012xr,Zimmerman:2016qtn,Godazgar:2017igz,Gralla:2017lto,Gralla:2018xzo,Hadar:2017ven,Hadar:2018izi}. It has been argued that the higher-order radial derivatives of the scalar field show the polynomial growth on the future Poincar\'e horizon. In their study (in fact, also in the original Aretakis{'}s study), the divergent behavior of the higher-order derivatives has been shown in specific coordinate systems. 
Thus, it is not trivial whether this divergent behavior is just a coordinate effect or not.
In the case of the extremal Reissner-Nordstr\"om black holes, there is a unique timelike Killing vector $V$ which is the generator of the event horizon. In the Eddington-Finkelstein coordinates $(v,r)$ where the timelike Killing vector~$V$ is a coordinate basis, the radial derivative operator $\partial_r$ satisfies ${\cal L}_{V} \partial_r = 0$. Then, the growth of some components of a tensor in the Eddington-Finkelstein coordinates is not a coordinate effect. Because the Aretakis instability, i.e., the growth of $\partial_r^{n} \Phi$ with an integer $n$, implies that $\nabla_r \nabla_r \cdots \nabla_r \Phi = \partial_r^{n} \Phi + {\rm lower~derivatives}$ is divergent, this is not a coordinate effect. 
However, in the case of $AdS_2$, there are many possible timelike Killing vectors, and there is no unique way to choose one of them. Actually, if we choose a coordinate system where one of the coordinate bases is the global timelike Killing vector, we can show that the higher-order derivatives do not blow up. Thus, one may think that the Aretakis instability in $AdS_2$ is due to the choice of the coordinate systems~\cite{Lucietti:2012xr, Hadar:2017ven}.\footnote{In Ref.~\cite{Hadar:2017ven}, there is an argument that the Aretakis instability in $AdS_2 \times S^2$ is not a coordinate effect if we consider $AdS_2 \times S^2$ as a near horizon geometry of extremal black holes. This is because the $AdS$ structure in the near horizon geometry appears in the Poincar\'e chart and the generator of the Poincar\'e horizon can be regarded as the horizon generator of the original black hole spacetime.} In this paper, to make this point clear, we revisit to study the Aretakis constants and instability in $AdS_2$. We find out the geometrical meaning of the Aretakis instability in the parallelly propagated (parallel-transported) null geodesics frame on the horizon, i.e., some components of the higher-order covariant derivatives of the field in the parallelly propagated frame blow up at the late time. In general relativity, parallelly propagated frames are used for studying the singular behavior of tensors in a coordinate independent way. For example, if the components of the Riemann tensor in the parallelly propagated frame are divergent at some point, we regard the point as a curvature singularity even if all scalar quantities constructed from the Riemann tensor, e.g., the Ricci scalar or the Kretschmann invariant, are finite~\cite{Hawking:1973uf}. Thus, our result implies the divergent behavior of the covariant derivatives of the fields at the late time. In the study of the Aretakis instability, the conserved quantities on the horizon, called the Aretakis constants, make the analysis easier~\cite{Aretakis:2011ha,Aretakis:2011hc,Aretakis:2012ei,Lucietti:2012xr,Lucietti:2012sf,Murata:2012ct,Gralla:2019isj}. In this paper, we also show that Aretakis constants in $AdS_2$ become some components of the higher-order covariant derivatives of the field in the parallelly propagated frame.

Because $AdS_2$ is maximally symmetric, any null hypersurfaces have the same geometrical properties. If we prepare the parallelly propagated null geodesic frame along any null hypersurfaces, the above discussion holds not only on the future Poincar\'e horizon but also on any null hypersurfaces. This implies that the Aretakis instability is the result of singular behaviors of the higher-order covariant derivatives of the fields on the whole $AdS$ infinity, rather than a blow-up on a specific null hypersurface. Also, by focusing on the maximal symmetry of $AdS_2$, we can construct scalar quantities that are constant not only on the future Poincar\'e horizon but also on any null hypersurfaces, and reduce to the Aretakis constants on the future Poincar\'e horizon. In this paper, we call these scalar quantities the generalized Aretakis constants. In Ref. \cite{Cardoso:2017qmj}, it has been shown that the ladder operators constructed from the spacetime conformal symmetry of $AdS_2$ lead to conserved quantities on any null hypersurfaces, and checked that they coincide with the generalized Aretakis constants for special mass squared cases. In this paper, we explicitly show the relation with the generalized  Aretakis constants for general cases. We also discuss that the generalized Aretakis constants and instability in $AdS_2$ are related to the conformal Killing tensors. 

This paper is organized as follows. In section \ref{sec:2}, we briefly review the Aretakis constants and instability in $AdS_2$ based on Ref. \cite{Lucietti:2012xr}. In section \ref{sec:3}, we introduce the parallelly propagated null geodesics frame, and discuss the Aretakis constants and instability in that frame. We also generalize to the case for any null hypersurfaces by using parallelly propagated frames on them. In section \ref{sec:4}, we discuss a relation between the generalization of the Aretakis constants and the spacetime conformal symmetry in $AdS_2$. In the final section, we summarize this paper. In Appendix \ref{appendix:massladder}, we review the mass ladder operators in $AdS_2$~\cite{Cardoso:2017qmj,Cardoso:2017egd}. Appendix \ref{appendix:relbtwnAQinAdS2} gives the proof of proposition.3 introduced in Sec. \ref{subsec:4-B}.

\section{The Aretakis constants and instability in $AdS_2$}
\label{sec:2}
We briefly review the Aretakis constants and instability in $AdS_2$ \cite{Aretakis:2011ha,Aretakis:2011hc,Aretakis:2012ei,Lucietti:2012xr,Zimmerman:2016qtn,Godazgar:2017igz,Gralla:2017lto,Gralla:2018xzo,Hadar:2017ven,Hadar:2018izi}. In the ingoing Eddington-Finkelstein coordinates $(v,r)$, $AdS_2$ is described by
\begin{equation}
\label{ads2}
ds^2=-r^2dv^2+2dvdr,
\end{equation}
where the future Poincar\'e horizon is located at $r=0$. We consider massive scalar fields $\Phi(v,r)$ in $AdS_2$. The fields obey the massive Klein-Gordon equation
\begin{equation}
\label{eom}
2\partial_v\partial_r\Phi+\partial_r\left(r^2\partial_r\Phi\right)-m^2\Phi=0.
\end{equation}

For mass squared $m^2=\ell(\ell+1)$ $(\ell=0,1,2,\cdots)$, acting the $\ell$-th order derivative operator $\partial_r^{\ell}$ on Eq. \eqref{eom} and evaluating it at $r=0$ show
\begin{equation}
\label{con1}
\left.\partial_v\partial_{r}^{\ell+1}\Phi\right|_{r=0}=0.
\end{equation}
This shows that $\mathcal{H}_\ell$ defined by
\begin{equation}
\label{arec}
\mathcal{H}_\ell:=\left.\partial_{r}^{\ell+1}\Phi\right|_{r=0},
\end{equation}
are independent of $v$. Hence, $\mathcal{H}_\ell$ are conserved quantities along the future Poincar\'e horizon, and then called \textit{the Aretakis constants} in $AdS_2$. For other mass squared, such conserved quantities on the future Poincar\'e horizon cannot be found. Differentiating Eq.~\eqref{eom} $(\ell+1)$ times with respect to $r$, we obtain
\begin{equation}
\left.\partial_v\partial_r^{\ell+2}\Phi\right|_{r=0}=-(\ell+1)\mathcal{H}_\ell.
\end{equation}
This implies
\begin{equation}
\label{arecinstability}
\partial_r^{\ell+2}\Phi|_{r=0}=-(\ell+1)\mathcal{H}_\ell v+{\rm const.}
\end{equation}
We see that $(\ell+2)$-th order derivative of the field on the future Poincar\'e horizon will blow up at the late time if $\mathcal{H}_\ell\neq0$. This divergent behavior is called \textit{the Aretakis instability} in $AdS_2$. We note that the $(\ell+3)$-th or higher-order derivatives are polynomially divergent at the late time.

For general mass squared cases with $m^2 \ge m_{\rm BF}^{2}=-1/4$, where $m_{\rm BF}^{2}$ is the Breitenlohner-Freedman bound~\cite{Breitenlohner:1982jf, Breitenlohner:1982bm} in $AdS_2$, Ref. \cite{Lucietti:2012xr} has shown that the late-time behavior of the $n$-th derivatives of the fields with respect to $r$ at the future Poincar\'e horizon $r=0$ becomes\footnote{Note that we focus on the normalizable modes.}
\begin{align}
\label{divergentbehavior}
\partial_r^{n}\Phi\big|_{r =0} \sim v^{n - \Delta_m},
\end{align}
where $n$ is a non-negative integer and
\begin{equation}
\label{Deltam}
\Delta_m :=\frac{1}{2} + \sqrt{m^2 + \frac{1}{4}}.
\end{equation}
Hence, using the notation,
\begin{equation}
\label{nm}
n_{m} := \lfloor{\Delta_{m}}\rfloor + 1,
\end{equation}
where $\lfloor{\Delta_{m}}\rfloor$ denotes the integer part of $\Delta_{m}$, the $n_{m}$-th order derivative of the field at $r = 0$ will blow up at the late time. This is also called the Aretakis instability in $AdS_2$. We note that the $(n_m+1)$-th or higher-order derivatives are also unbounded.

\section{The Aretakis constants and instability in the parallelly propagated null geodesics frame}
\label{sec:3}
In this section, we discuss the geometrical meaning of the Aretakis constants and instability in the parallelly propagated null geodesic frame. We shall show that some components of the higher-order covariant derivatives of the field in the parallelly propagated frame are constant or unbounded, and they correspond to the Aretakis constants and instability, respectively.

\subsection{On the future Poincar\'e horizon}
\label{sec:Poincarehorizon}
We first discuss the late-time divergent behavior in Eq. \eqref{divergentbehavior}. We introduce vector fields on the future Poincar\'e horizon $r=0$,
\begin{align}
e_{(0)}^\mu \partial_\mu = \partial_v,~e_{(1)}^\mu \partial_\mu = -\partial_r,
\end{align}
where these satisfy 
\begin{equation}
\begin{split}
\label{basisrelation1AdS}
& e_{(0)}^\mu \nabla_\mu e_{(0)}^\nu = 0,~ 
e_{(0)}^\mu \nabla_\mu e_{(1)}^\nu = 0, 
\\
& e_{(0)}^\mu e_{(0)\mu} = 0,~ 
e_{(1)}^\mu e_{(1)\mu} = 0,~ 
e_{(0)}^\mu e_{(1)\mu} = -1,
\end{split}
\end{equation}
at $r=0$. Hence, $e_{(1)}^\mu$ is parallelly transported along the null geodesic $e_{(0)}^\mu$ on the future Poincar\'e horizon. The frame formed by $(e_{(0)}^\mu,e_{(1)}^\mu)$ is called \textit{the parallelly propagated null geodesic frame} on the future Poincar\'e horizon.

For the massive scalar $\Phi(v,r)$ satisfying Eq. \eqref{eom} with general mass squared and positive integer $n$, the following relation holds:
\begin{equation}
\begin{split}
\label{ppdivergent}
&\left.\left(-1\right)^{n}e_{(1)}^{\mu_1}e_{(1)}^{\mu_2} \cdots e_{(1)}^{\mu_{n}}
\nabla_{\mu_1}\nabla_{\mu_2} \cdots \nabla_{\mu_{n}} \Phi\right|_{r=0}
= \left.\partial_{r}^{n} \Phi\right|_{r=0}.
\end{split}
\end{equation}
Using the notation $n_m$ is defined in Eq. \eqref{nm}, the divergent behavior of $\partial_{r}^{n_m} \Phi$ at $r  = 0$ implies that the $n_m$-th order covariant derivative is also divergent in the parallelly propagated null geodesic frame. We note that for $n\le n_m$, the components of $\nabla_{\mu_1}\cdots\nabla_{\mu_{n}} \Phi$
in the parallelly propagated null geodesic frame
are bounded except for the $e_{(1)}^{\mu_1}e_{(1)}^{\mu_2} \cdots e_{(1)}^{\mu_{n}}$ component with $n=n_m$ from Eq.~\eqref{divergentbehavior}.\footnote{For positive integers $n$ and $q$, the following relation holds:
\begin{equation}
\begin{split}
\label{e0components}
&\left(-1\right)^{n}\left.
e_{(0)}^{\mu_1} \cdots e_{(0)}^{\mu_{q}}
e_{(1)}^{\nu_{1}} \cdots e_{(1)}^{\nu_{n}}
\nabla_{\mu_1}\cdots \nabla_{\mu_{q}} 
\nabla_{\nu_{1}} \cdots \nabla_{\nu_{n}} 
\Phi\right|_{r = 0}= \left.\partial_v^{q} \partial_{r}^{n} \Phi\right|_{r = 0}.
\end{split}
\end{equation}
Other components of the covariant derivatives in the parallelly propagated frame can be written by Eq.~\eqref{e0components} and the lower order derivatives using the commutation relation for the covariant derivatives.}

For the mass squared $m^2=\ell(\ell+1)$ $(\ell=0,1,2,\cdots)$, we obtain
\begin{equation}
\begin{split}
\label{AretakisinPP}
\left.\left(-1\right)^{\ell+1}e_{(1)}^{\mu_1}e_{(1)}^{\mu_2} \cdots e_{(1)}^{\mu_{\ell+1}}
\nabla_{\mu_1}\nabla_{\mu_2} \cdots \nabla_{\mu_{\ell+1}} \Phi\right|_{r=0}=\left.\partial_{r}^{\ell+1} \Phi\right|_{r = 0}.
\end{split}
\end{equation}
We find that $e_{(1)}^{\mu_1}e_{(1)}^{\mu_2} \cdots e_{(1)}^{\mu_{\ell+1}}$ component of the $(\ell+1)$-th order covariant derivative of the field on the future Poincar\'e horizon is the Aretakis constant $\mathcal{H}_\ell$ in Eq. \eqref{arec}.

\begin{figure}
  \centering
   \includegraphics[width=7cm]{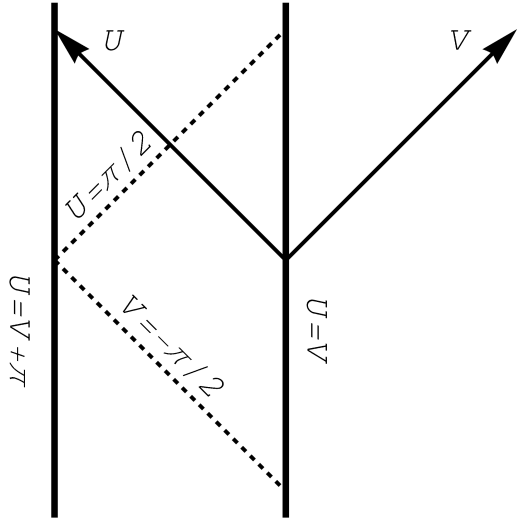}
  \caption{The Penrose diagram for $AdS_2$. The left and right $AdS$ boundaries are located at $U=V+\pi$ and $U=V$, respectively. The future and past Poincar\'e horizons are located at $U=\pi/2$ and $V=-\pi/2$, respectively.}\label{PenrosediagramAdS2}
\end{figure}

\subsection{On any null hypersurfaces}
\label{sec:nullhypersurfaces}
Because $AdS_2$ is maximally symmetric,  the discussion in the previous subsection should also hold for any other null hypersurfaces. This implies that the Aretakis instability is the result of singular behaviors of the higher-order covariant derivatives of the fields on the whole $AdS$ infinity, rather than a blow-up on a specific null hypersurface. In this subsection, we explicitly show that for $m^2 = \ell (\ell + 1)$ $(\ell=0,1,2,\cdots)$ cases.

\subsubsection{The massive scalar fields in the global chart $(U,V)$}
For the latter convenience, we discuss the massive scalar fields in the global chart \cite{Cardoso:2017qmj}. In the double null global chart $(U,V)$ defined by
\begin{equation}
\label{UVads2}
\tan U=v+\frac{2}{r},~\tan V=v,
\end{equation}
the line element in Eq. \eqref{ads2}, which describes $AdS_2$, is rewritten as
\begin{equation}
ds^2=-\frac{4}{H(U,V)}dUdV,
\end{equation}
where
\begin{equation}
\label{HUV}
H(U,V)=\sin^2(U - V).
\end{equation}
The coordinate range is 
$-\infty < U < \infty, -\infty < V < \infty$ with $0 < U - V < \pi$,
and the $AdS$ boundary locates at $V = U$ or $V = U + \pi$ where $H(U,V)=0$. The future and past Poincar\'e horizons are $U=\pi/2$ and $V=-\pi/2$, respectively. The Penrose diagram of $AdS_2$ is shown in FIG.~\ref{PenrosediagramAdS2}.

In the present coordinates, the massive Klein-Gordon equation \eqref{eom} is rewritten as
\begin{equation}
\label{eomUV}
\left[-H(U,V)\partial_V\partial_U-m^2\right]\Phi(U,V)=0,
\end{equation}
where $H(U,V)$ is given by Eq. \eqref{HUV}. We notice that for the massless scalar, this equation shows $\partial_V\partial_U\Phi=0$, and hence $\partial_U\Phi$ is constant along any null hypersurfaces $U={\rm const}$. At the future Poincar\'e horizon $U=\pi/2$, it coincides with the Aretakis constant $\mathcal{H}_0$ in Eq.~\eqref{arec}. Thus, $\partial_U\Phi$ is the generalization of the Aretakis constant $\mathcal{H}_0$. According to Ref. \cite{Cardoso:2017qmj}, for the mass squared $m^2=\ell(\ell+1)$, there exists the generalization of the Aretakis constants $\mathcal{H}_\ell$ in Eq. \eqref{arec} for general $\ell$,\footnote{If we exchange $U$ and $V$ in Eq. \eqref{garec}, we can construct conserved quantities on $V={\rm const.}$ surfaces.}
\begin{equation}
\label{garec}
\mathcal{A}_\ell:=\left(\frac{\cos^2V}{H(U,V)}\right)^{\ell+1}\left[\frac{H(U,V)}{\cos^2V}\partial_U\right]^{\ell+1}\Phi,
\end{equation}
and they satisfy
\begin{equation}
\partial_v\mathcal{A}_\ell=0.
\end{equation}
We call $\mathcal{A}_\ell$ \textit{the generalized Aretakis constants}. It is easy to check that $\mathcal{A}_\ell=\mathcal{H}_\ell$ at the future Poincar\'e horizon $U=\pi/2$. For other mass squared cases, conserved quantities on a null hypersurface cannot be found.

\subsubsection{The parallelly propagated null geodesic frame in $AdS_2$}
Now, we introduce null vector fields
\begin{equation}
\begin{split}
\label{basisads}
\mathfrak{e}_{(0)}^\mu \partial_\mu = \frac{f(U)H(U,V)}{4}\partial_{V},~
\mathfrak{e}_{(1)}^\mu \partial_\mu = \frac{2}{f(U)}\partial_{U},
\end{split}
\end{equation}
where $f(U)$ is an arbitrary finite function. These satisfy the relations 
\begin{equation}
\begin{split}
\label{relas1AdS}
& \mathfrak{e}_{(0)}^\mu \nabla_\mu \mathfrak{e}_{(0)}^\nu = 0,~ \mathfrak{e}_{(0)}^\mu \nabla_\mu \mathfrak{e}_{(1)}^\nu = 0, 
\\
& \mathfrak{e}_{(0)}^\mu \mathfrak{e}_{(0)\mu} = 0,~ \mathfrak{e}_{(1)}^\mu\mathfrak{e}_{(1)\mu} = 0,~ \mathfrak{e}_{(0)}^\mu \mathfrak{e}_{(1)\mu} = -1.
\end{split}
\end{equation}
Therefore, $(\mathfrak{e}_{(0)}^\mu, \mathfrak{e}_{(1)}^\mu)$ form the parallelly propagated null geodesic frame for each null hypersurface $U={\rm const.}$ We should note that $\mathfrak{e}^{\mu}_{(0)}$ and $\mathfrak{e}^{\mu}_{(1)}$ are vector fields defined in the whole $AdS_2$ spacetime, while $e_{(0)}^\mu$ and $ e_{(1)}^\mu$ in Eq.~\eqref{basisrelation1AdS} are defined only on $r = 0$ surface. Hereafter, we set $f(U) = 2$. We note that this specific choice of $f(U)$ does not change the conclusion in the following discussions.

\subsubsection{Massless scalar cases}
\label{sec:massless}
General solutions of the massless Klein-Gordon equation \eqref{eomUV} are
\begin{equation}
\Phi(U,V) =F(U) + G(V).
\end{equation}
Then, the generalized Aretakis constant in Eq. \eqref{garec} is $\mathcal{A}_0=\partial_UF(U)$. Now, we can see
\begin{align}
\mathfrak{e}^{\mu}_{(1)} \nabla_\mu \Phi &= \mathcal{A}_0,\label{aretakisuvcoordads2}
\\
\label{e1e1chi}
\mathfrak{e}^{\mu}_{(1)}\mathfrak{e}^{\nu}_{(1)} \nabla_\mu\nabla_\nu \Phi &= 
\mathcal{A}_0\frac{\partial_U H(U,V)}{H(U,V)}+ \partial_U^{2}F(U).
\end{align}
Eq. \eqref{aretakisuvcoordads2} shows that the geometrical meaning of ${\mathcal A}_0$ at each null hypersurface $U = {\rm const.}$ is the same as the Aretakis constant ${\cal H}_0$ at $r = 0$, i.e., a component of the covariant derivative in the parallelly propagated frame is constant at each null hypersurface. Because $\partial_U H/H=2/\tan(U-V)$, $\mathfrak{e}^{\mu}_{(1)}\mathfrak{e}^{\nu}_{(1)} \nabla_\mu\nabla_\nu \Phi$ in Eq. (\ref{e1e1chi}) is divergent linearly in $(U-V)^{-1}$ at the $AdS$ boundary if $\mathcal{A}_0 \neq 0$.\footnote{If we consider the ``normalizable" mode $\Phi \sim (U-V)^{\Delta_m}$ with $\Delta_m = 1$, where $\Delta_m$ is defined by Eq. \eqref{Deltam}, for the massless case at the $AdS$ boundary $U = V$, then the function $G(V)$ becomes $G(V) = - F(V)$.
We note that the Aretakis instability occurs even in that case.} Near the $AdS$ boundary, we further show
\begin{equation}
\label{e1e1Kchi}
\mathfrak{e}^{\mu_{1}}_{(1)}\mathfrak{e}^{\mu_{2}}_{(1)}\cdots\mathfrak{e}^{\mu_{n+2}}_{(1)} \nabla_{\mu_{1}}\nabla_{\mu_{2}}\cdots\nabla_{\mu_{n+2}} \Phi=\mathcal{A}_0\mathcal{O}\left((U-V)^{-n-1}\right),
\end{equation}
where $n\ge1$. Eqs. \eqref{e1e1chi} and \eqref{e1e1Kchi} show that the second and higher-order covariant derivatives of the field \textit{on any null hypersurfaces} have singular behaviors at the $AdS$ boundary if $\mathcal{A}_0\neq0$.\footnote{If $\mathcal{A}_0=0$ and $\partial_U^2F(U)\neq0$, the third-order derivative $\mathfrak{e}^{\mu_1}_{(1)}\mathfrak{e}^{\mu_2}_{(1)}\mathfrak{e}^{\mu_3}_{(1)} \nabla_{\mu_1}\nabla_{\mu_2}\nabla_{\mu_3} \Phi $ is divergent (see proposition.1 in Sec. \ref{sec:proposition}). } We comment that other components are bounded,
 \begin{align}
\label{e0chi}
\mathfrak{e}^{\mu}_{(0)} \nabla_\mu \Phi &= \frac{H(U,V)}{2}\partial_VG(V) ,\\
\mathfrak{e}^{\mu}_{(0)}\mathfrak{e}^{\nu}_{(0)} \nabla_\mu\nabla_\nu \Phi &= 
\frac{ H(U,V)}{4}\partial_V\left(H(U,V)\partial_VG(V)\right),\\
\mathfrak{e}^{\mu}_{(0)}\mathfrak{e}^{\nu}_{(1)} \nabla_\mu\nabla_\nu \Phi &= 
\mathfrak{e}^{\mu}_{(1)}\mathfrak{e}^{\nu}_{(0)} \nabla_\mu\nabla_\nu \Phi= 0.
\label{e0e1chi}
\end{align}

\subsubsection{Massive scalar cases with $m^2 = \ell(\ell +1)$}
For the cases $m^2 = \ell(\ell +1)$ $(\ell=1,2,\cdots)$, we can also explicitly see the divergent behavior at the $AdS$ boundary.
For $\ell = 1$ case, the general normalizable Klein-Gordon fields, which are derived in Appendix \ref{appnedix:massladderUV}, take the form of
\begin{equation}
\begin{split}
\Phi(U,V) =& \frac{2 \cos U \cos V}{\sin(U-V)}\left(F(U) - F(V)\right)-  \cos^2 U \partial_UF(U) -\cos^2 V \partial_VF(V),
\label{l1kgfield}
\end{split}
\end{equation}
with an arbitrary function $F$.\footnote{
Note that $\Phi\left(U,V\right)$ satisfies the normalizable boundary condition, i.e., $\Phi \sim (U-V)^{\Delta_m}$
with $\Delta_m = \ell+ 1$, at $V = U$.
If we also impose this condition at $V = U + \pi$,
$F$ should satisfy $F(U) = F(U + \pi)$.
}
We obtain
\begin{equation}
\begin{split}
\label{e1e1chi1}
&\mathfrak{e}^{\mu_1}_{(1)}\mathfrak{e}^{\mu_2}_{(1)} \nabla_{\mu_1}\nabla_{\mu_2} \Phi
=\mathcal{A}_1,
\end{split}
\end{equation}
where $\mathcal{A}_1=2(-1 + 2 \cos(2U))\partial_UF(U)+\cos U \big(6\sin U\partial_U^{2}F(U)   
-  \cos U\partial_U^{3}F(U) \big)$ is the generalized Aretakis constant \eqref{garec} and  
\begin{equation}
\begin{split}
\label{e1e1e1chi}
\mathfrak{e}^{\mu_1}_{(1)}\mathfrak{e}^{\mu_2}_{(1)} \mathfrak{e}^{\mu_3}_{(1)}
\nabla_{\mu_1}\nabla_{\mu_2} \nabla_{\mu_3} \Phi=2\mathcal{A}_1\frac{\partial_U H(U,V)}{H(U,V)}+\partial_U\mathcal{A}_1.
\end{split}
\end{equation}
Because $\partial_U H/H=2/\tan(U-V)$, Eq. \eqref{e1e1e1chi} shows that the third-order covariant derivative of the field on any null hypersurfaces has the linear growth of $(U-V)^{-1}$ at the $AdS$ boundary if $\mathcal{A}_1\neq0$. We comment that other components are bounded. 

For $\ell \ge 2$, acting the mass ladder operators, which are given by Eq. \eqref{massladderinUV}, we can easily obtain the explicit form of the general normalizable Klein-Gordon field, which is Eq.~\eqref{massrasingPhi} with $G(V)=-F(V)$. We can show that  $(\ell+1)$-th and $(\ell+2)$-th order covariant derivatives are, respectively, constant along each null hypersurface and divergent at the $AdS$ boundary, 
\begin{equation}
\begin{split}
\label{e1e1garec}
\mathfrak{e}^{\mu_1}_{(1)}\mathfrak{e}^{\mu_2}_{(1)} \cdots\mathfrak{e}^{\mu_{\ell+1}}_{(1)}\nabla_{\mu_1}\nabla_{\mu_2} \cdots\nabla_{\mu_{\ell+1}} \Phi=\mathcal{A}_\ell,
\end{split}
\end{equation}
where $\mathcal{A}_\ell$ is the generalized Aretakis constant in Eq. \eqref{garec} and
\begin{equation}
\begin{split}
\label{e1e1e1divergent}
\mathfrak{e}^{\mu_1}_{(1)}\mathfrak{e}^{\mu_2}_{(1)} \cdots\mathfrak{e}^{\mu_{\ell+2}}_{(1)}\nabla_{\mu_1}\nabla_{\mu_2} \cdots\nabla_{\mu_{\ell+2}} \Phi=(\ell+1)\mathcal{A}_\ell\frac{\partial_U H(U,V)}{H(U,V)}+\partial_U\mathcal{A}_\ell.
\end{split}
\end{equation}

\subsection{Relation between the conserved quantities on the null hypersurface and divergent behavior}
\label{sec:proposition}
We have observed that, for a solution of the massive Klein-Gordon equation \eqref{eomUV} with the mass squared $m^2 = \ell (\ell +1)$ $(\ell=0,1,2,\cdots)$ in $AdS_2$, in the parallelly propagated null geodesic frame, $(\ell+1)$-th covariant derivative of the field gives a constant along each null hypersurface and $(\ell+2)$-th covariant derivative has a linear divergent behavior along the null hypersurface.
We can generalize this relation, i.e., the relation between 
a conserved quantity on a null hypersurface and the divergent behavior, as follows:
\\
\\
{\bf Proposition.1~}
\textit{
If the relation
\begin{align}
\mathfrak{e}^{\mu_{n}}_{(1)} \mathfrak{e}^{\mu_{n-1}}_{(1)}\cdots\mathfrak{e}^{\mu_1}_{(1)}  \nabla_{\mu_{n}}  \nabla_{\mu_{n-1}}\cdots \nabla_{\mu_1}  \Psi = A(U),
\label{eq:aretakisconst_a}
\end{align}
holds for some scalar field $\Psi(U,V)$ in $AdS_2$, a positive integer $n$, and a regular function $A(U) (\not \equiv 0)$,
then
$\mathfrak{e}^{\mu_{n+1}}_{(1)}\mathfrak{e}^{\mu_n}_{(1)}\cdots\mathfrak{e}^{\mu_{1}}_{(1)}  \nabla_{\mu_{n+1}}  \nabla_{\mu_n}\cdots \nabla_{\mu_{1}}  \Psi$ is divergent at the $AdS$ boundary.
}
\\
\\
{\bf Proof}.~~
Acting an operator $\mathfrak{e}^{\mu_{n+1}}_{(1)} \nabla_{\mu_{n+1}}$ to Eq.~\eqref{eq:aretakisconst_a},
we obtain 
\begin{align}
\mathfrak{e}^{\mu_{n+1}}_{(1)}\mathfrak{e}^{\mu_n}_{(1)}\cdots\mathfrak{e}^{\mu_{1}}_{(1)}  \nabla_{\mu_{n+1}}  \nabla_{\mu_n}\cdots \nabla_{\mu_{1}}  \Psi=&\partial_UA(U)
-\mathfrak{e}^{\mu_{n+1}}_{(1)} \nabla_{\mu_{n+1}}\left(\mathfrak{e}^{\mu_{n}}_{(1)} \cdots\mathfrak{e}^{\mu_1}_{(1)}\right)  \nabla_{\mu_{n}}  \cdots \nabla_{\mu_1} \Psi\notag\\
=&\partial_UA(U)+nA(U)\frac{\partial_U H(U,V)}{H(U,V)},
\label{omega1omega1ddpsi}
\end{align}
where we have used a relation
\begin{align}
& \mathfrak{e}^{\mu}_{(1)}\nabla_\mu \mathfrak{e}^{\nu}_{(1)}
=-\frac{\partial_U H(U,V)}{H(U,V)}\mathfrak{e}^{\nu}_{(1)}.
\label{omega1eq}
\end{align}
In the right-hand side of Eq.~\eqref{omega1omega1ddpsi},
the first term is finite but the second term is divergent at the $AdS$ boundary $V = U$ (or $V = U +\pi$) because $\partial_U H/H=2/\tan(U-V)$. \hfill$\Box$
$~$
\\

If $\Psi(U,V)$ is the massive Klein-Gordon field with the mass squared $m^2 = \ell (\ell + 1)$, the above proposition leads to the relation between the generalized Aretakis constant and the divergent behavior at the $AdS$ boundary.

As another application of the above proposition with $n = 1$, for the massive Klein-Gordon fields $\Phi(U,V)$ with the mass squared $m^2 = \ell (\ell +1)$ in $AdS_2$,
if we choose the function $\Psi(U,V)$ as 
\begin{align}
\label{masslessPsi}
\Psi = D_{i_{1},1} D_{i_{2},2} \cdots D_{i_\ell,\ell} \Phi,
\end{align}
where $D_{i,\ell}$ are the mass ladder operators in Eq. \eqref{massladderinUV}, then $\Psi(U,V)$ satisfies the massless Klein-Gordon equation \eqref{eomUV}. Then, $\mathfrak{e}^{\mu}_{(1)} \nabla_\mu \Psi\propto \partial_U\Psi$ is a constant along each null hypersurface, which corresponds 
to the generalized Aretakis constant $\mathcal{A_\ell}$ in Eq. \eqref{garec} as will be shown in section \ref{sec:4},
and $\mathfrak{e}^{\mu}_{(1)} \mathfrak{e}^{\nu}_{(1)} \nabla_\mu \nabla_\nu \Psi$ is linearly 
divergent along the null hypersurface.

We can also show the following proposition:
~~\\
~~\\
{\bf Proposition.2~}
\textit{ If the relation
\begin{align}
\mathfrak{e}^{\mu_{n}}_{(1)} \mathfrak{e}^{\mu_{n-1}}_{(1)}\cdots\mathfrak{e}^{\mu_1}_{(1)}  \nabla_{\mu_{n}}  \nabla_{\mu_{n-1}}\cdots \nabla_{\mu_1}  \Psi  = A_0 + A_1(V)(U-U_0) + {\cal O}((U-U_0)^2),
\label{eq:aretakisconst_b}
\end{align}
holds for some scalar field $\Psi(U,V)$ in $AdS_2$, a positive integer $n$, a constant $A_0$ ($\neq 0$), and a bounded function $A_1(V)$,
then
$\mathfrak{e}^{\mu_{n+1}}_{(1)}\mathfrak{e}^{\mu_n}_{(1)}\cdots\mathfrak{e}^{\mu_{1}}_{(1)}  \nabla_{\mu_{n+1}}  \nabla_{\mu_n}\cdots \nabla_{\mu_{1}}  \Psi$ 
is divergent at the $AdS$ boundary along $U= U_0$.
}
\\
\\
{\bf Proof}.~~
If we set $A = A_0 + A_1(V)(U-U_0) + {\cal O}((U-U_0)^2)$, Eq.~\eqref{omega1omega1ddpsi} still holds.
Because $A_1$ is a bounded function, 
$\mathfrak{e}^{\mu_{n+1}}_{(1)}\mathfrak{e}^{\mu_n}_{(1)}\cdots\mathfrak{e}^{\mu_{1}}_{(1)}  \nabla_{\mu_{n+1}}  \nabla_{\mu_n}\cdots \nabla_{\mu_{1}}  \Psi$ is divergent at the $AdS$ boundary along $U= U_0$. \hfill$\Box$
$~$
\\

We note that scalar fields $\Psi(U,V)$ in the above propositions are not necessarily the massive Klein-Gordon fields. Proposition.2 shows that the existence of a constant along a null hypersurface leads to the divergent behavior of the higher derivative. Finally, we comment that proposition.2 holds if $A_0$ is a function of $V$ and has a non-vanishing limiting value $\lim_{V\to \infty} A_0 \neq 0$.

\subsection{The relation among the conformal Killing tensors, the Aretakis constants and instability }
\label{sec:conformalKilling}
For positive integers $n$, rank-$n$ tensors
\begin{equation}
\label{conformalKillingtensor}
K^{\mu_1 \mu_2 \cdots \mu_{n}} :=\mathfrak{e}_{(1)}^{\mu_1}\mathfrak{e}_{(1)}^{\mu_2}\cdots \mathfrak{e}_{(1)}^{\mu_{n}},
\end{equation}
are conformal Killing tensors in $AdS_2$ and the only non-trivial components are $K^{UU\cdots U} = 1$.\footnote{We note that $K^{\mu_1 \mu_2 \cdots \mu_{n}}$ is parallelly propagated along $\mathfrak{e}_{(0)}^{\mu}$, i.e., $\mathfrak{e}_{(0)}^{\nu}\nabla_\nu K^{\mu_1 \mu_2 \cdots \mu_{n}}=0$, and satisfies ${\cal L}_\xi K^{\mu_1 \mu_2 \cdots \mu_n} = 0$ with the Killing vector $\xi = \partial_V + \partial_U$.}
For the scalar fields $\Phi(U,V)$ with the mass squared $m^2=\ell(\ell+1)$ $(\ell=0,1,2,\cdots)$, Eq. \eqref{e1e1garec} shows that the generalized Aretakis constants $\mathcal{A}_\ell$ in Eq. \eqref{garec} relate with the rank-$(\ell+1)$ conformal Killing tensor \cite{Cardoso:2017qmj},
\begin{equation}
K^{\mu_1 \mu_2 \cdots \mu_{n_{\ell+1}}}\nabla_{\mu_1}\nabla_{\mu_2} \cdots \nabla_{\mu_{\ell+1}} \Phi
=\mathcal{A}_\ell.
\end{equation}
Eq. \eqref{e1e1e1divergent} implies near the $AdS$ boundary $V\simeq U$,
\begin{equation}
\label{KDPhi}
K^{\mu_1 \mu_2 \cdots \mu_{n_{\ell+2}}}\nabla_{\mu_1}\nabla_{\mu_2} \cdots \nabla_{\mu_{\ell+2}} \Phi=2(\ell+1)\frac{\mathcal{A}_\ell}{U-V}+\mathcal{O}\left((U-V)^0\right).
\end{equation}
Hence, the contraction with the rank-$(\ell+2)$ conformal Killing tensor and the $(\ell+2)$-th order covariant derivative will blow up linearly in $(U-V)^{-1}$ at the $AdS$ boundary if $\mathcal{A}_\ell\neq 0$. 

For the general mass squared $m^2\ge m^2_{\rm BF}=-1/4$, where the Aretakis constants do not necessarily exist, we have the relation
\begin{equation}
K^{\mu_1 \mu_2 \cdots \mu_{n_m}}\nabla_{\mu_1}\nabla_{\mu_2} \cdots \nabla_{\mu_{n_m}} \Phi= \mathfrak{e}_{(1)}^{\mu_1}\mathfrak{e}_{(1)}^{\mu_2}\cdots \mathfrak{e}_{(1)}^{\mu_{n_m}}\nabla_{\mu_1}\nabla_{\mu_2} \cdots \nabla_{\mu_{n_m}} \Phi,
\end{equation}
where the notation $n_m$ is defined in Eq. \eqref{nm}. As discussed in Secs. \ref{sec:Poincarehorizon} and \ref{sec:nullhypersurfaces}, the right-hand side is divergent at the $AdS$ boundary. Thus, the Aretakis instability can also be regarded as that the contraction with the conformal Killing tensor $K^{\mu_1 \mu_2 \cdots \mu_{n_m}}$ and the $n_m$-th order covariant derivative of the Klein-Gordon field is divergent at the $AdS$ boundary.

\section{The Aretakis constants from the spacetime conformal symmetry}
\label{sec:4}
In this section, we discuss the relation between the generalized Aretakis constants in $AdS_2$ in Eq.~\eqref{garec} and the ladder operators  constructed from the spacetime conformal symmetry~\cite{Cardoso:2017qmj,Cardoso:2017egd} for massive Klein-Gordon fields with the mass squared $m^2=\ell(\ell+1)$ $(\ell=0,1,2,\cdots)$. First, we construct conserved quantities at each null hypersurface $U = {\rm const.}$ following~Ref. \cite{Cardoso:2017qmj}. Next, we show that they coincide with the generalized Aretakis constants up to constant factors. Note that cases for $\ell=1,2$ have been discussed~\cite{Cardoso:2017qmj}. 

\subsection{Conserved quantities at each null hypersurface from the mass ladder operators}
\label{subsec:4-A}
We discuss the scalar fields $\Phi(U,V)$ obeying the massive Klein-Gordon equation \eqref{eomUV} with the mass squared $m^2=\ell(\ell+1)$. First, let us consider the massless case $\ell = 0$. The massless Klein-Gordon equation \eqref{eomUV} shows
\begin{equation}
\partial_V\partial_U\Phi=0.
\end{equation}
We can see that $\partial_U\Phi$ is a conserved quantity at each null hypersurface $U={\rm const.}$, and this quantity is the generalized Aretakis constant $\mathcal{A}_0$ in Eq. \eqref{garec}.

Next, we consider $\ell \ge 1$ cases. Using the mass ladder operators \cite{Cardoso:2017qmj,Cardoso:2017egd} (see Appendix~\ref{appendix:massladder} for a brief review), the massive Klein-Gordon fields can be mapped into the massless Klein-Gordon fields. Following Ref.~\cite{Cardoso:2017qmj}, we can construct conserved quantities at each null hypersurface $U = {\rm const.}$ similar to the massless case. The explicit calculation is shown below. From the relation \eqref{eq:comrelformassladderinAdS2} with $k=s=\ell$ on the scalar field~$\Phi$, we obtain
\begin{equation}
\begin{split}
\label{massladderell}
D_{i_{\ell},-1}D_{i_{\ell-1},0}\cdots D_{i_{1},\ell-2}\left[-H(U,V)\partial_V\partial_U-\ell(\ell+1)\right]\Phi=-H(U,V)\partial_V\partial_U D_{i_\ell,1}D_{i_{\ell-1},2}\cdots D_{i_{1},\ell}\Phi,
\end{split}
\end{equation}
where the mass ladder operators $D_{i,k}$ are given by Eq.~\eqref{massladderinUV} and $H(U,V)$ is given by Eq.~\eqref{HUV}. Since the left-hand side vanishes due to the Klein-Gordon equation for $\Phi$, Eq.~\eqref{massladderell} leads to
\begin{equation}
\begin{split}
\label{massless_eom}
-H(U,V)\partial_V\partial_U D_{i_\ell,1}D_{i_{\ell-1},2}\cdots D_{i_{1},\ell}\Phi=0.
\end{split}
\end{equation}
Thus, solutions of the massive Klein-Gordon equation with the mass squared $m^2 = \ell(\ell +1)$ in $AdS_2$ can be mapped into that of the massless Klein-Gordon equation. We note that massive fields with other mass squared cannot be mapped into massless fields.
As in the case $\ell=0$, Eq.~\eqref{massless_eom} shows
\begin{equation}
\label{conservationlawUV}
\partial_V \mathcal{Q}_\ell = 0,
\end{equation}
where
\begin{align}
\label{QinUV}
{\cal Q}_\ell := W(U)\partial_U D_{i_\ell,1}D_{i_{\ell-1},2}\cdots D_{i_{1},\ell}\Phi.
\end{align}
For later convenience, using $\partial_V W(U) = 0$, we have added an arbitrary function $W(U)$ as a factor. Eq.~\eqref{conservationlawUV} shows that ${\cal Q}_\ell$ are conserved quantities at each null hypersurface $U={\rm const.}$ As will be discussed below, the quantity ${\cal Q}_\ell $ relates to the generalized Aretakis constant $\mathcal{A}_\ell$. 

\subsection{The relation with the Aretakis constants on the future Poincar\'e horizon}
\label{subsec:4-B}
We shall show that ${\cal Q}_\ell$ coincide with the Aretakis constants $\mathcal{H}_\ell$ in Eq. \eqref{arec} on the future Poincar\'e horizon $U=\pi/2$ by choosing $W(U)$ appropriately. It is convenient to use the ingoing Eddington-Finkelstein coordinates $(v,r)$. Using $\partial_U=-(2+2vr+(1/2+v^2/2)r^2)\partial_r$, Eq. \eqref{QinUV} is written as
\begin{align}
{\cal Q}_\ell= -W(U)\left(2+2vr+\frac{1+v^2}{2}r^2\right)\partial_r D_{i_\ell,1}D_{i_{\ell-1},2}\cdots D_{i_{1},\ell}\Phi.
\end{align}
Because $\tan{U}=v+2/r$ in Eq. \eqref{UVads2}, we can regard $W$ as a function of $v+2/r$. Hereafter, we consider $W = -2^{-1}C_W(v/2 + 1/r)^q$ cases, where $C_W$ and $q$ are constants. Then, we can evaluate the leading term of ${\cal Q}_\ell$ as 
\begin{align}
\label{QAdS}
{\cal Q}_\ell= C_Wr^{-q}\partial_r D_{i_\ell,1}D_{i_{\ell-1},2}\cdots D_{i_{1},\ell}\Phi\left(1+\mathcal{O}\left(r\right)\right).
\end{align}
By choosing $C_W$ and $q$ appropriately, we can show that ${\cal Q}_\ell $ coincide with the Aretakis constants on the future Poincar\'e horizon, $\mathcal{H}_\ell$ in Eq. \eqref{arec}. For this purpose, we introduce the following proposition.
~~\\
~~\\
{\bf Proposition.3~}
\textit{For analytic solutions of the massive Klein-Gordon equation with the mass squared $\ell (\ell + 1), ~(\ell = 0, 1, 2, \cdots)$ in $AdS_2$,
\begin{align}
\left[2\partial_v\partial_r+2r\partial_r+r^2\partial_r^{2} - \ell(\ell + 1)\right]\Phi(v,r) = 0,
\label{eq:AdS2kgeq}
\end{align}
the relation
\begin{equation}
\begin{split}
&2^{- n_{1}+n_{-1}} r^{-2 n_{-1}- n_0} \partial_r D_{i_\ell,1}D_{i_{\ell-1},2}\cdots D_{i_{1},\ell}\Phi=  \partial_r^{\ell+1} \Phi + {\cal O}(r),
\label{eq:d1eqinads2}
\end{split}
\end{equation}
holds, where $n_{-1}, n_0, n_{1}$ are the numbers of the mass ladder operators 
constructed from $\zeta_{-1}, \zeta_{0}, \zeta_1$, respectively, included in the left-hand side of Eq. \eqref{eq:d1eqinads2}.
The numbers $n_{-1}, n_0, n_{1}$ satisfy $n_{-1}+ n_0+ n_{1}= \ell$.
}
~~\\
~~\\
The proof is given in Appendix \ref{appendix:relbtwnAQinAdS2}. Because $\mathcal{H}_\ell=\partial_r^{\ell+1}\Phi|_{r=0}$, the above proposition and Eq.~\eqref{QAdS} show $\mathcal{Q}_\ell|_{r=0}=\mathcal{H}_\ell$ if $C_W=2^{- n_{1}+n_{-1}}$ and $q=2 n_{-1}+ n_0$. We should note that regardless of the choice of the closed conformal Killing vectors $\zeta_{-1}, \zeta_{0}, \zeta_1$, $\mathcal{Q}_\ell|_{r=0}$ relate with the same conserved quantities $\mathcal{H}_\ell$.

\subsection{The relation with the generalized Aretakis constants}
\label{subsec:4-C}
Next, we shall discuss that ${\cal Q}_\ell$ in Eq. \eqref{QinUV} coincide with the generalized Aretakis constants $\mathcal{A}_\ell$ in Eq. \eqref{garec} by choosing $W(U)$ appropriately. In the construction of $Q_\ell$ in Eq.~\eqref{QinUV}, if we replace the mass ladder operators $D_{i,k}$ with the general mass ladder operators in Eq.~\eqref{generalmlo}, $\mathcal{Q}_\ell$ are still independent of $V$. In that case, if all general mass ladder operators contain $\zeta_1$, $\mathcal{Q}_\ell$ at the Poincar\'e horizon coincide with $\mathcal{Q}_\ell$ constructed only from $\zeta_1$ up to the constant factor because of proposition.3. 
Because $AdS_2$ is maximally symmetric, we can generalize this to other null hypersurfaces $U = {\rm const.}$, i.e., if all general closed conformal Killing vectors to construct $\mathcal{Q}_\ell$ are not proportional to $\partial_V$ at a null hypersurface $U = U_0$, then those $\mathcal{Q}_\ell$ at $U = U_0$ are proportional to $\mathfrak{e}^{\mu_1}_{(1)}\mathfrak{e}^{\mu_2}_{(1)} \cdots\mathfrak{e}^{\mu_{\ell+1}}_{(1)}\nabla_{\mu_1}\nabla_{\mu_2} \cdots\nabla_{\mu_{\ell+1}} \Phi$. Thus, $\mathcal{Q}_\ell$ in Eq.~\eqref{QinUV} coincide with the generalized Aretakis constant ${\mathcal A}_\ell$ up to the factor of a function of $U$ because $\zeta_{-1}, \zeta_0, \zeta_1$ are not proportional to $\partial_V$ except at the Poincar\'e horizon.

\section{Summary and discussion}
In this paper, we have studied the geometrical meaning of the 
Aretakis constants and instability for massive scalar fields in $AdS_2$.
We have shown that the Aretakis constants and instability in $AdS_2$ can be understood as that
some components of the higher-order covariant derivatives of the scalar fields in the parallelly propagated null geodesic frame
are constant or unbounded at the future Poincar\'e horizon.
Due to the maximal symmetry of $AdS_2$, the same discussion holds not only on the future Poincar\'e horizon but also on any null hypersurfaces. 
We then have clarified that the generalization of the Aretakis constants~\cite{Cardoso:2017qmj} called \textit{the generalized Aretakis constants} have the same geometrical meaning as that in the future Poincar\'e horizon, i.e., 
some components of the higher-order covariant derivatives 
in the parallelly propagated null geodesic frame
are constant at each null hypersurface.
Also, we have seen that the higher-order covariant derivatives of the scalar fields have singular behaviors at the whole $AdS$ boundary, and that causes the Aretakis instability in $AdS_2$.
If we consider cases for the mass squared with $m^2_{\rm BF} < m^2 < 0$, where 
$m^2_{\rm BF} = -1/4$ is the Breitenlohner-Freedman (BF) bound~\cite{Breitenlohner:1982jf, Breitenlohner:1982bm}, 
the first order covariant derivatives of the scalar fields are divergent at the $AdS$ boundary. This implies that some physical quantities such as the energy-momentum tensor have also divergent behaviors at the $AdS$ boundary for $m^2_{\rm BF} < m^2 < 0$.

We have also discussed the relation with the spacetime conformal symmetry.
For the fields with the mass squared $m^2=\ell(\ell+1)$ $(\ell=0,1,2,\cdots)$, the contraction with the rank-$(\ell+2)$ conformal Killing tensor and the $(\ell+2)$-th order covariant derivatives of the field is divergent at the whole $AdS$ boundary if the generalized Aretakis constant exists. 
If we see this divergent behavior on a null hypersurface, it corresponds to the Aretakis instability.
We note that the generalized Aretakis constants can be expressed as the contraction with the rank-$(\ell+1)$ conformal Killing tensor and the $(\ell+1)$-th order covariant derivatives~\cite{Cardoso:2017qmj}.
We have demonstrated that the generalized Aretakis constants can be derived from the mass ladder operators constructed from the closed conformal Killing vectors~\cite{Cardoso:2017qmj}. 

Since the $AdS_2$ structures appear in the vicinity of extremal black hole horizons \cite{Kunduri:2013gce,Lucietti:2012xr,Zimmerman:2016qtn,Godazgar:2017igz,Gralla:2017lto,Gralla:2018xzo,Hadar:2017ven,Hadar:2018izi}, we expect that the Aretakis instability in extremal black hole spacetimes has a similar geometrical meaning as our result in $AdS_2$ cases. In fact, this expectation is correct, i.e., the Aretakis instability for black hole cases \cite{Aretakis:2011ha,Aretakis:2011hc,Aretakis:2012ei,Lucietti:2012sf,Murata:2012ct,Gralla:2019isj} can be understood as that some components of the higher-order covariant derivatives of the field in the parallelly propagated frame are unbounded at the late time \cite{arecinBH}.

\begin{acknowledgments}
The authors would like to thank Shahar Hadar, Tomohiro Harada, Takaaki Ishii, Shunichiro Kinoshita, Keiju Murata, Shin Nakamura, Harvey S. Reall, and Norihiro Tanahashi for useful comments and discussions. M.K. acknowledges support by MEXT Grant-in-Aid for Scientific Research on Innovative Areas 20H04746.
\end{acknowledgments}

\appendix

\section{The mass ladder operators in $AdS_2$}
\label{appendix:massladder}
We briefly review the mass ladder operators \cite{Cardoso:2017qmj, Cardoso:2017egd} in $AdS_2$,
which map solutions of the massive Klein-Gordon equation into that with the different mass squared.

\subsection{Spacetime conformal symmetries and mass ladder operators}
\label{subsubsec:4-A}
It is said that an $n$-dimensional spacetime $\left(\mathcal{M}, g_{\mu\nu}\right)$ possesses a spacetime conformal symmetry if the metric $g_{\mu\nu}$ admits a conformal isometry $\phi$ defined by $\phi: \mathcal{M}\to \mathcal{M}$ such that $\phi^*g_{\mu\nu}=\exp{\left(2Q\right)} g_{\mu\nu}$, where $Q$ is a function on $\mathcal{M}$. The transformation of the conformal isometry group is generated by an infinitesimal coordinate transformation $x^\mu\to \bar{x}^\mu=x^\mu-\zeta^\mu$ along a vector field $\zeta^\mu$ called a \textit{conformal Killing vector}. The conformal Killing vector $\zeta^\mu$ obeys the \textit{conformal Killing equation}
\begin{equation}
\label{cke}
\mathcal{L}_\zeta g_{\mu\nu}=2Qg_{\mu\nu},~~Q=\frac{1}{n}\nabla_{\mu}\zeta^\mu.
\end{equation} 
A conformal Killing vector is said to be \textit{closed} if $\nabla_{[\mu}\zeta_{\nu]}=0$ is satisfied. Then, the closed conformal Killing vector satisfies the \textit{closed conformal Killing equation}
\begin{equation}
\label{ccke}
\nabla_{\mu}\zeta_\nu=Qg_{\mu\nu}.
\end{equation}
If the closed conformal Killing vector $\zeta^\mu$ further satisfies
\begin{equation}
\label{ricci}
R^\mu_{~\nu}\zeta^\nu=\lambda~\zeta^\mu,
\end{equation}
where $\lambda$ is a constant, we can define \textit{the mass ladder operator} \cite{Cardoso:2017qmj,Cardoso:2017egd},
\begin{equation}
\label{mlo}
D_{k}:=\mathcal{L}_\zeta-kQ,~~k\in\mathbb{R},
\end{equation}
which maps a solution of the massive Klein-Gordon equation to that with a different mass squared, i.e., for a solution of the massive Klein-Gordon equation,
\begin{equation}
\label{eommassive}
\left[\Box-m^2\right]\Phi=0,
\end{equation}
with $m^2 = -\lambda k(k+n-1)$, $D_{k}\Phi$ becomes another solution of the massive Klein-Gordon equation
\begin{equation}
[\square - (m^2+ \delta m^2)]D_{k}\Phi = 0,
\label{eq:dkphi}
\end{equation}
with $m^2+ \delta m^2 = -\lambda (k-1)(k+n-2)$.
Here, we have used the commutation relations for $D_{k}$,
\begin{equation}
[\square, D_{k}] = \lambda (2k + n-2) D_k + \frac{2}{n}(\nabla_\mu \zeta^\mu)\left[\square + \lambda k (k+n-1)\right].
\label{eq:comrelformassladder}
\end{equation}
Note that the condition \eqref{ricci} is automatically satisfied for vacuum solutions of the Einstein equations with a cosmological constant, e.g., the Anti-de Sitter spacetime. For a given mass squared with $(n-1)^2 - 4m^2/\lambda \ge 0$ ($\lambda\neq0$), there are two possible $k = k_\pm$ as solutions of $m^2 = -\lambda k(k+n-1)$, 
\begin{equation}
\label{kpm}
k_\pm = \frac{1-n \pm \sqrt{(n-1)^2 - 4 m^2/\lambda}}{2}.
\end{equation}
The mass ladder operators $D_{k}$ correspond to a mass raising or lowering operator, depending on the sign of $\lambda$.

If there exist two or more closed conformal Killing vectors, we can investigate the Lie bracket among them. It is defined by
\begin{equation}
\label{kc}
\xi_{~i,j}^{\mu}:=\left[\zeta_i,\zeta_j\right]^\mu=\zeta_{~i}^{\nu}\nabla_\nu \zeta_{~j}^{\mu}-\zeta_{~j}^{\nu}\nabla_\nu \zeta_{~i}^{\mu},
\end{equation}
where the indices ``$i$" and``$ j$" label the different closed conformal Killing vectors. Then, the vectors \eqref{kc} satisfy the Killing equation $\mathcal{L}_\xi g_{\mu\nu}=0$, where we have used Eq. \eqref{ricci}.

\subsection{The mass ladder operators in $AdS_2$}
\label{subsubsec:4-A-2}
In the $AdS_2$ cases, i.e., $n=2$ and $\lambda = -1$, the mass ladder operators exist when $m^2 \ge m^2_{\rm BF} = -1/4$. Note that this condition corresponds to the non-negativity of the inside of the square root in Eq.~\eqref{kpm}. For the massive Klein-Gordon equation \eqref{eommassive} in $AdS_2$, $k_{\pm}$ in Eq. \eqref{kpm} are 
\begin{equation}
\label{kpminAdS}
k_{+}=\nu,~~k_{-}=-(\nu+1),
\end{equation}
where we parameterized the mass squared as $m^2=\nu(\nu+1)$.\footnote{In the derivation of Eq.~\eqref{kpminAdS}, we have assumed $\nu \ge -1/2$. If $\nu<-1/2$, $k_+ = -(\nu+1)$ and $k_- = \nu$. We note that $m^2 = \nu(\nu+1)$ with $\nu \ge -1/2$ corresponds to $m^2 \ge m^2_{\rm BF}$, thus it is enough to consider $\nu\ge -1/2$ cases.} Noted that $k_-=-\Delta_{m}$ in Eq.~\eqref{Deltam}.

Solving the closed conformal Killing equation \eqref{ccke} for $AdS_2$, we obtain three closed conformal Killing vectors,
\begin{equation}
\begin{split}
\label{ckvads}
\zeta_{-1}&=\partial_v+r^2\partial_r,\\
\zeta_{0}&=v\partial_v+\left(r +vr^2\right)\partial_r,\\
\zeta_{1}&=v^2\partial_v+\left(2+2v r+v^2r^2\right)\partial_r.
\end{split}
\end{equation}
We here note that $\zeta^\mu_{-1}$ and $\zeta^\mu_{0}$ become null on the Poincar\'e horizon, while $\zeta^\mu_{1}$ does not. We comment that the Lie bracket (\ref{kc}) among the closed conformal Killing vectors  $\zeta_i^\mu$ ($i=-1,0,1$) yields three Killing vectors,
\begin{equation}
\begin{split}
\label{kv}
\xi_{-1}&\left(:=\xi_{0,-1}\right)=\partial_v,\\
\xi_{0}&\left(:=\xi_{-1,1}\right)=v\partial_v-r\partial_r,\\
\xi_{1}&\left(:=\xi_{1,0}\right)=v^2\partial_v-2\left(1+vr\right)\partial_r.
\end{split}
\end{equation}

Using the closed conformal Killing vectors \eqref{ckvads}, we obtain three mass ladder operators,
\begin{equation}
\begin{split}
\label{massladderinAdS}
D_{-1,k}=&\partial_v+r^2\partial_r-kr,\\
D_{0,k}=&v\partial_v+r\left(1+v r\right)\partial_r-k\left(1+v r\right),\\
D_{1,k}=&v^2\partial_v+\left(2+2v r+v^2r^2\right)\partial_r-kv\left(2+v r\right).
\end{split}
\end{equation}
For the closed conformal Killing vectors $\zeta_i^\mu$ in Eq. (\ref{ckvads}), the mass ladder operators $D_{i,k}$ defined in Eq. \eqref{massladderinAdS} map a solution of the massive Klein-Gordon equation \eqref{eommassive} with $m^2 = k(k+1)$ to that with a different mass squared $m^2 + \delta m^2 =(k-1)k$. Note that if we consider the general closed conformal Killing vectors $\zeta = a_{-1} \zeta_{-1} + a_0 \zeta_0 + a_1 \zeta_1$, where $a_{-1},a_{0},a_{1}$ are constants, we can construct the general mass ladder operators in $AdS_2$ as 
\begin{align}
\label{generalmlo}
D_k = a_{-1} D_{-1,k} + a_0 D_{0,k} + a_1  D_{1,k}.
\end{align}

The commutation relation \eqref{eq:comrelformassladder} can be written as
\begin{equation}
D_{i,k-2}\left[\Box-k(k+1)\right]=\left[\square-k(k-1)\right]D_{i,k}.
\end{equation}
Using this, for a positive integer $s$, we can show
\begin{equation}
\label{eq:comrelformassladderinAdS2}
D_{i_s,k-s-1}\cdots D_{i_2,k-3}D_{i_1,k-2}\left[\Box-k(k+1)\right]=\left[\square-(k-s)(k+s-1)\right]D_{i_s,k-s+1}\cdots D_{i_2,k-1}D_{i_1,k}.
\end{equation}

\subsection{The mass ladder operators in the global chart}
\label{appnedix:massladderUV}
We here introduce the mass ladder operators in the $(U,V)$ chart in Eq.~\eqref{UVads2},
\begin{equation}
\begin{split}
\label{massladderinUV}
D_{-1,k}=&\cos^2V\partial_V-\cos^2U\partial_U-k\frac{2\cos V\cos U}{\sin\left(U-V\right)},\\
D_{0,k}=&\sin V\cos V\partial_V-\sin U\cos U\partial_U-k\frac{\sin\left(U+V\right)}{\sin\left(U-V\right)},\\
D_{1,k}=&\sin^2 V\partial_V-\sin^2 U\partial_U-k\frac{2\sin U\sin V}{\sin\left(U-V\right)}.
\end{split}
\end{equation}
We note that the above mass ladder operators are regular differential operators except at the $AdS$ boundary, and the divergent behavior at the $AdS$ boundary changes the asymptotic behavior of the scalar fields near the $AdS$ boundary from $\Phi(U,V) \sim c_1 (U-V)^{-k} + c_2(U-V)^{(k+1)}$ to $D_{i,k}\Phi \sim c_1 (U-V)^{-(k-1)} + c_2(U-V)^{k}$,
where $c_1=c_1(U)$ and $c_2=c_2(U)$, and we have assumed the mass squared of the massive Klein-Gordon equation is $m^2 = k(k+1) (\ge -1/4)$
 ~\cite{Cardoso:2017qmj}.

We can construct the general solutions of the massive Klein-Gordon equation \eqref{eomUV} with the mass squared $m^2 = \ell(\ell + 1)~ (\ell = 1, 2, \cdots)$, from the general solution of the massless Klein-Gordon equation, $\Phi_0(U,V) = F(U) + G(V)$ as follows:
\begin{align}
\label{massrasingPhi}
\Phi_\ell(U,V)= D_{i_\ell,-\ell} D_{i_{\ell-1},-(\ell-1)} \cdots D_{i_1, -1} \Phi_0.
\end{align}
Because the mass ladder operators are surjective (onto) maps as shown in~\cite{Cardoso:2017qmj}, $\Phi_\ell(U,V)$ becomes the general solution of the massive Klein-Gordon equation.
For example, $\Phi_1(U,V)$ with $D_{-1,k}$ becomes
\begin{align}
\Phi_1(U,V)=\frac{2 \cos U \cos V}{\sin(U-V)}\left(F(U) + G(V)\right)-  \cos^2 U\partial_UF(U) + \cos^2 V\partial_VG(V).
\end{align}
If we impose the normalizable boundary condition at $U = V$, we obtain Eq.~\eqref{l1kgfield}.

\section{Proof of proposition.3}
\label{appendix:relbtwnAQinAdS2}
In this proof, for scalar fields with the mass squared $m^2=\ell(\ell+1)$ $(\ell=0,1,2,\cdots)$, we write $\Phi, n_{-1}, n_0, n_1$ as $\Phi_{\ell}, n_{-1}^\ell, n_0^\ell, n_1^\ell$, respectively. We expand $\Phi_\ell(v,r)$ as a Taylor series around $r = 0$,
\begin{align}
\label{TaylorPhi}
\Phi_\ell(v,r) = \sum_{s = 0}^\infty C_s^\ell(v) r^{s}.
\end{align}
where $C_s^\ell(v)$ is given by
$C_s^\ell(v) = (s!)^{-1} \left.\partial_r^{s} \Phi_\ell \right|_{r = 0}. $
The massive Klein-Gordon equation \eqref{eq:AdS2kgeq} becomes
\begin{align}
\sum_{s = 1}^{\infty}
\left[
2 s \frac{dC_s^\ell}{dv} + (s + \ell)(s - \ell - 1) C^\ell_{s-1}
\right] r^s = 0,
\end{align}
then, we obtain the relation\footnote{Note that the relation Eq.~\eqref{eq:dcmdv} implies 
$dC_{\ell+1}^k/dv = 0$, then $(\ell+1)!C_{\ell+1}^\ell = \mathcal{H}_\ell = {\rm const.}$ and
 $dC_{\ell+2}^\ell/dv = - C^\ell_{\ell + 1}(\ell + 1)/(\ell + 2)$, then $(\ell+2)! C_{\ell+2}^\ell = \partial_r^{\ell + 2} \Phi|_{r = 0}
= - (\ell+1)\mathcal{H}_\ell v+{\rm const.}$ These correspond to the Aretakis constants and instability in Eqs. \eqref{arec} and \eqref{arecinstability}.
We note that the coefficients $C^\ell_{s}$ with $s\le \ell$ are decaying functions of $v$ if we choose the normalizable modes.}

\begin{align}
 \frac{dC_s^\ell}{dv} = -  \frac{(s + \ell)(s - \ell - 1)}{2s} C^\ell_{s-1}.
\label{eq:dcmdv}
\end{align}
We would like to show that if the relation~\eqref{eq:d1eqinads2} holds for $\ell$, then it also holds for $\ell + 1$,
\begin{equation}
\begin{split}
&2^{- n_{1}^{\ell+1}+n_{0}^{\ell+1}} r^{-2 n_{-1}^{\ell+1} - n_0^{\ell+1}}\partial_r D_{i_\ell,1}D_{i_{\ell-1},2}\cdots D_{i_{1},\ell}
D_{i_{\ell+1},\ell+1}
\Phi_{\ell+1} =  \partial_r^{\ell+2} \Phi_{\ell+1} + {\cal O}(r),
\label{eq:d1eqinads2ellp1}
\end{split}
\end{equation}
where $n_{-1}^{\ell+1} + n_0^{\ell+1}+ n_{1}^{\ell+1} = \ell + 1$.
We note that relation~\eqref{eq:d1eqinads2} trivially holds for $\ell = 0$.
Substituting Eq.~\eqref{eq:dcmdv} into Eq.~\eqref{TaylorPhi}, after some straightforward calculations, 
we can show the relations
\begin{align}
&2 r^{-2} \partial_r^{\ell + 1} D_{-1,\ell+1} \Phi_{\ell + 1}
=
\partial_{r}^{\ell + 2} \Phi_{\ell + 1} + {\cal O}(r),
\\
&r^{-1} \partial_r^{\ell + 1} D_{0,\ell+1} \Phi_{\ell + 1}
=
\partial_{r}^{\ell + 2} \Phi_{\ell + 1} + {\cal O}(r),
\\
&2^{-1} \partial_r^{\ell + 1} D_{1,\ell+1} \Phi_{\ell + 1}
=
\partial_r^{\ell + 2} \Phi_{\ell+1} + {\cal O}(r).
\end{align}
These relations immediately lead to Eq.~\eqref{eq:d1eqinads2ellp1}.
As an example, we show the $i_{\ell+1} = -1$ case below.
Since $D_{i_{\ell+1},\ell+1}\Phi_{\ell + 1}$ is a solution of the Klein-Gordon equation with the mass squared $\ell(\ell + 1)$, 
we can set
\begin{align}
\Phi_\ell = D_{-1,\ell+1} \Phi_{\ell + 1}.
\end{align}
The left-hand side of Eq.~\eqref{eq:d1eqinads2ellp1} becomes
\begin{align}
& 2^{- n_{1}^{\ell}+ n_{-1}^{\ell} + 1} r^{-2 (n_{-1}^{\ell}+1) - n_0^{\ell}} \partial_r D_{i_\ell,1}D_{i_{\ell-1},2}\cdots D_{i_{1},\ell}
D_{-1,\ell+1}
\Phi_{\ell+1} 
\notag\\&=
2^{- n_{1}^{\ell}+ n_{-1}^{\ell} + 1}   r^{-2 (n_{-1}^{\ell}+1) - n_0^{\ell}} \partial_r D_{i_\ell,1}D_{i_{\ell-1},2}\cdots D_{i_{1},\ell}
\Phi_{\ell} 
\notag\\&=
2r^{-2} \partial_r^{\ell + 1} \Phi_\ell + {\cal O}(r)
\notag\\&=
2r^{-2} \partial_r^{\ell+ 1} D_{-1,\ell+1} \Phi_{\ell+ 1} + {\cal O}(r)
\notag\\&= 
\partial_{r}^{\ell+ 2} \Phi_{\ell + 1}  + {\cal O}(r).
\end{align}
Note that the cases for $i_{\ell + 1} = 0, 1$ can be shown in a same way.

\end{document}